\documentclass{PoS}

\PoS{PoS(HEP2005)277}

\usepackage{graphicx,cite,amssymb,epsfig,float,psfrag,multirow,amsmath,subfigure}

\catcode`@=11

\def\@citex[#1]#2{\if@filesw\immediate\write\@auxout{\string\citation{#2}}\fi
  \def\@citea{}\@cite{\@for\@citeb:=#2\do
    {\@citea\def\@citea{,\penalty\@m}\@ifundefined
      {b@\@citeb}{{\bf ?}\@warning
       {Citation `\@citeb' on page \thepage \space undefined}}%
\hbox{\csname b@\@citeb\endcsname}}}{#1}}

\def\citer{\@ifnextchar [{\@tempswatrue\@citexr}{\@tempswafalse\@citexr[]}}

\def\@citexr[#1]#2{\if@filesw\immediate\write\@auxout{\string\citation{#2}}\fi
  \def\@citea{}\@cite{\@for\@citeb:=#2\do
    {\@citea\def\@citea{--\penalty\@m}\@ifundefined
       {b@\@citeb}{{\bf ?}\@warning
       {Citation `\@citeb' on page \thepage \space undefined}}%
\hbox{\csname b@\@citeb\endcsname}}}{#1}}

\def\citer{\@ifnextchar [{\@tempswatrue\@citexr}{\@tempswafalse\@citexr[]}}
\catcode`@=12

\newcommand{\beq}{\begin{eqnarray}}
\newcommand{\eeq}{\end{eqnarray}}

\newcommand{\nn}{\nonumber}
\newcommand{\lsim}{\stackrel{<}{_\sim}}

\def\gsim{ {\ \lower-1.2pt\vbox{\hbox{\rlap{$>$}\lower5pt\vbox{\hbox{$\sim$}}}}\ } }
\title{Phenomenological study of the double radiative decay 
$B \to K \gamma \gamma$}

\ShortTitle{Phenomenological study of the double radiative decay 
$B \to K \gamma \gamma$}

\vspace*{-0.4cm}
\author{Gudrun Hiller\\
    Institut f\"ur Physik, Universit\"at Dortmund, D-44221 Dortmund, Germany\\
        E-mail: \email{ghiller@physik.uni-dortmund.de}}

\vspace*{-0.4cm}
\author{\speaker{A.~Salim Safir}
\\
    CERN, Department of Physics, Theory Unit, CH-1211 Geneva 23, Switzerland\\
        E-mail: \email{safir@mail.cern.ch}}

\abstract{
Using the operator product expansion (OPE) technique, we study the rare double
radiative decay $B\to K \gamma\gamma$ in the Standard Model (SM) and beyond. We
estimate the short-distance (SD) contribution to the decay amplitude in a
region of the phase space which is around the point where all decay products
have energy $\sim m_b/3$ in the rest frame of the $B$-meson.
At lowest order in $1/m_b$, the $B\to K \gamma\gamma$ matrix element is then
expressed in terms of the usual $B\to K$ form factors known from semileptonic
rare decays.
The integrated SD branching ratio in the SM in the OPE region turns out to be
$\Delta {\cal{B}}(B \to K \gamma \gamma)_{SM}^{OPE} \simeq 1 \times 10^{-9}$.
We work out the di-photon invariant mass distribution with and without the
resonant background through $B\to K \{\eta_c,\chi_{c0}\}\to K\gamma \gamma$. In
the SM, the resonance contribution is dominant in the region of phase space
where the OPE is valid. On the other hand, the present experimental upper limit
on $B_s \to \tau^+ \tau^-$ decays, leaves considerable room for New Physics 
(NP) in
the one-particle-irreducible contribution to $B\to K \gamma \gamma$ decays. In
this case, we find
that the SD $B\to K \gamma \gamma$ branching ratio can be enhanced by one order
of magnitude with respect to its SM value and the SD contribution can lie
outside of the resonance peaks.}

\FullConference{International Europhysics Conference on High Energy Physics\\

                 July 21st - 27th 2005\\

                 Lisboa, Portugal}

\begin{document}
\begin{figure}[h]
\begin{center}
\epsfig{file=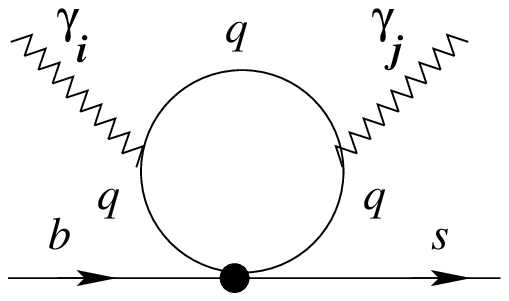,width=4cm,height=2.2cm}
\hspace*{2.cm}\epsfig{file=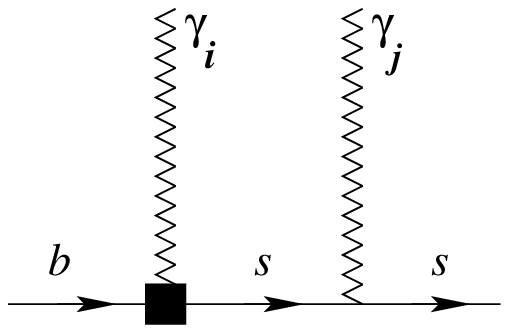,width=4cm,height=2.2cm}
\hspace*{0.5cm}\epsfig{file=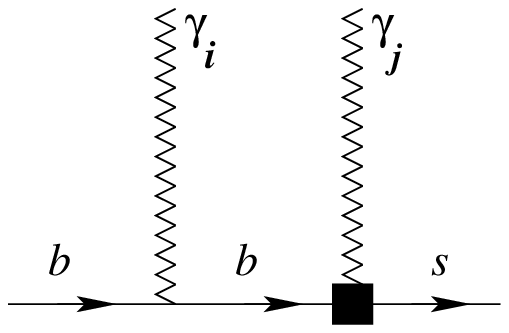,width=4cm,height=2.2cm}
\end{center}
\caption{\it Leading order Feynman diagrams for $b\to s \gamma\gamma$ decays.
Left plot: 1PI diagram from four-fermion operators, other two plots:
insertion of $O_7$. Figure taken from \cite{HS}.
\label{fig1}}
\end{figure}

%
Flavor changing neutral currents (FCNC) $B$-decays are of utmost importance 
in exploring and shaping the flavor structure 
of the SM and its extensions, since they are forbidden in the 
Born approximation and highly sensitive to NP contributions in the loops.
Although many studies, either experimentally 
or theoretically, have been devoted to 
$b\to s \gamma$ and $b\to s \ell^+\ell^-$ transitions, less 
attention has been paid to the  
double-photon decays  mediated by $b \to s \gamma \gamma$.
At the quark level
it receives one-particle-reducible (1PR) 
contributions from the $b \to s \gamma $ transition plus an additional 
photon and a one-particle-irreducible (1PI) term from a fermion loop with 
the two photons emitted from that loop (see Fig~\ref{fig1}). 
At the lowest order, the corresponding amplitude can be written 
as \cite{HS,Reina:1997my} 
\begin{eqnarray}
{\cal{A}}(b \to s \gamma \gamma) =
-\frac{i e^2 G_F}{\sqrt{2} \pi^2} \lambda_t
\bar{u}_{s}(p^\prime)  \cdot \left [ \frac{4}{9}(3 C_1 +C_2)
 \kappa_c W_2^{\mu \nu} \right. 
-\left. \frac{1}{3} C_7  W_7^{\mu \nu} \right ] \cdot  u_b(p) 
\epsilon_{\mu}(k_1) \epsilon_{\nu}(k_2),~~~~~~~
\end{eqnarray}
where  
$\lambda_t=V_{t b} V_{t s}^*$ and
$p (p^\prime)$ represents the momentum of the $b(s)$-quark. We
denote by $k_1,k_2$ the 4-momenta and by $\epsilon(k_{1,2})$ 
the polarization vectors of the photons.
The tensors   $W_{2}^{\mu \nu}$ and $W_{7}^{\mu \nu}$  
exhibit the contribution of the four-fermion operators ${\cal{O}}_{1,2} \simeq
( \bar s \gamma_{\mu} L c) (\bar c \gamma^\mu L b)$
and the photon dipole operator ${\cal{O}}_7 \sim \bar s \sigma_{\mu \nu} F^{\mu \nu}R b$, respectively, and are 
given together with the loop function $\kappa_c$ in \cite{HS}.

In this work we consider the 
exclusive $B \to K \gamma \gamma$ decay.
Since its matrix element 
induced by ${\cal{O}}_7$ is non-local, it is
difficult to estimate model-independently. Previous works 
\cite{Singer:1997ti,Choudhury:2002yu} used here solely 
vector meson dominance models.
Instead, we use OPE techniques, which allow to {\it i} make the
matrix element local in a specific kinematic region by choosing
appropriate photon energies and then {\it ii}
express the matrix element in terms of known form factors.
Specifically, we demand both the internal $s$- and $b$-quarks to be far 
off-shell, with virtualities of order $m_b$.
For example, if both
photons have energies $m_b/3$, then 
the intermediate propagators of the 1PR diagrams 
$(Q_{1,2}^s)^2=(p^\prime+k_{2,1})^2$ and
$(Q_{1,2}^b)^2=- ((p-k_{1,2})^2-m_b^2)$
equal  $m_b/\sqrt{3}$ and
$m_b \sqrt{2/3}$, respectively.
Then we integrate out large scales of order $m_b$.
We construct the vertices in the effective theory 
out of a bottom heavy HQET quark 
$h_v$ and a strange collinear SCET quark $\chi$ \cite{Bauer:2000yr,Beneke:2002ph}.
Here,  $v=p_B/m_B$ and a light-like vector $n=p_K/E_K$, 
where $E_K, p_K$ denotes the energy, 4-momentum of the kaon.
Hence, we perform an OPE in $\Lambda_{QCD}/Q$,
where $Q=\{ m_b, E_K, Q_{1,2}^{s,b},\sqrt{q^2} \}$ 
and $q^2=(k_1+k_2)^2$.

For the lowest order 
matching onto the $b \to s \gamma \gamma$ amplitude, 
we obtain after using the
equations of motion the following dimension 8 operators \cite{HS}
\begin{eqnarray}
\frac{m_b}{4}\bar \chi \sigma_{\mu \nu}\sigma_{\alpha \beta} R h_v 
F_{1}^{\alpha \beta} F_{2}^{\mu \nu}, ~~~~
-2 i m_b \bar \chi \sigma_{\mu \nu} R h_v F_{1}^{\mu \nu} 
F_{2}^{\alpha \beta} v^\alpha n^\beta, ~~~~~
\bar \chi \gamma^\mu L  h_v F_{1}^{\alpha \beta} D_\alpha \tilde F_{2 \, \beta \mu} ~~~ + (1 \leftrightarrow 2) \nn
\end{eqnarray} 
where
$\tilde F_{\mu \nu} =1/2 \epsilon_{\mu \nu \alpha \beta} F^{\alpha \beta}$ and
Wilson lines are understood in $\chi$.
The $B \to K \gamma \gamma$ matrix elements 
are then obtained from
tree level matching of the QCD onto the SCET currents 
\cite{Bauer:2000yr,Charles:1998dr,Beneke:2000wa} as 
\begin{align}
\langle K(n)|\bar \chi h_v | B(v) \rangle & = 2 E_K \zeta(E_K), &
\langle K(n)|\bar \chi\sigma_{\mu \nu} h_v | B(v) \rangle & = 
-2 i E_K \zeta(E_K) \left( v_\mu n_\nu -v_\nu n_\mu \right),\nn\\
\langle K(n)|\bar \chi \gamma_\mu h_v |B(v) \rangle & = 
2 E_K \zeta(E_K) n_\mu, &
\langle K(n)|\bar \chi \sigma_{\mu \nu} \gamma_5 h_v | B(v) \rangle & =
- 2 E_K \zeta(E_K) \epsilon_{\mu \nu \alpha \beta} v^\alpha n^\beta,
\end{align}
where the form factor $\zeta(E_K)$ can be identified with the QCD form factor 
in the usual parametrization, as $\zeta=f_+$. It is the only form factor
remaining in the symmetry limit, known e.g.~from studies in
$B \to K \ell^+ \ell^-$ decays. Analytical formulae  for the 
$B \to K \gamma \gamma$ amplitude from the OPE are given in \cite{HS}.
%
%


An additional contribution to $B\to K\gamma\gamma$  beyond the 
OPE is photon radiation off the spectator. It is either
kinematically excluded (soft gluons between active quarks and spectator) 
or suppressed by $\alpha_s$. Annihilation contributions are leading power
in $1/Q$,
but suppressed by CKM elements or small penguin Wilson coefficients 
(for further details, see \cite{HS}). 


Long-distance effects from $B \to (\eta_x \to \gamma \gamma) K$, 
where $\eta_x=\eta, \eta^\prime, \eta_c$ and
$B \to (K^* \to K \gamma) \gamma$ are sizeable in $B \to K \gamma \gamma$ 
decays \cite{Singer:1997ti,Choudhury:2002yu}.
Both the $K^*$- and the $\eta^{(\prime)}$-channels require kinematics that
is outside of our OPE window.
Therefore, we take in our analysis  
only the charmonia, specifically the $\eta_c$ and also the 
$\chi_{c0}$, into account.
(The current upper bound on ${\cal{B}}(B \to \chi_{c2} K)$ implies that the
background from the $\chi_{c2}$ 
is at least one order of magnitude down with respect to the scalar one.) 
We estimate the long-distance contamination from $\eta_c,\chi_{c0}$ with a
Breit-Wigner ansatz fitted to data \cite{HS}.

\begin{figure}[t]
\vskip 0.0truein
\begin{center}
\includegraphics[height=12.cm,width=8.cm,angle=270]{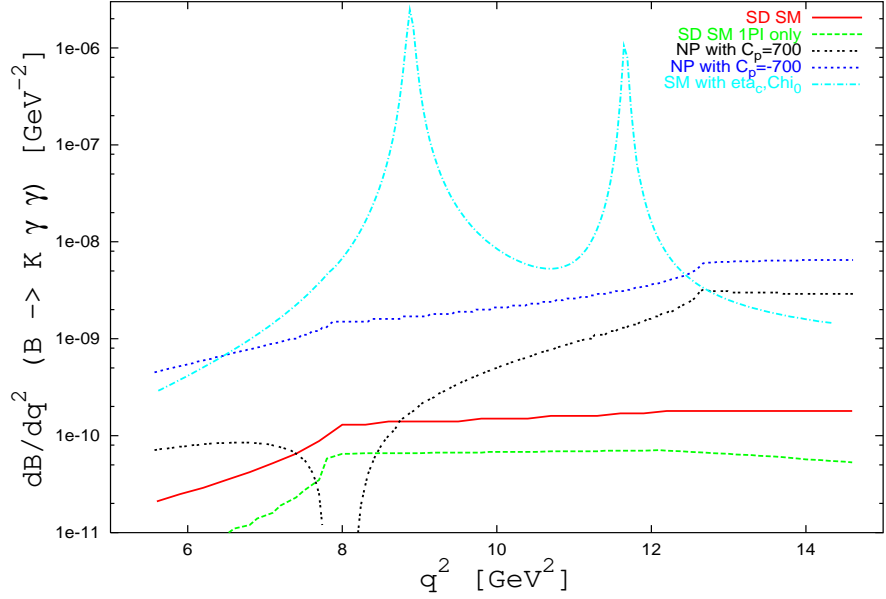}
\vskip 0.0truein
\end{center}
\caption[]{\it Di-photon mass distribution of 
$B \to K \gamma \gamma$ decays in the OPE region. The solid (dashed) curve denote the SM pure SD contribution with (and without) the 1PR terms. The  dotted (double-dotted)
curve corresponds to a NP scenario with $ C^\tau_{P}=-700 (+700)$ and
$C^\tau_{S}=0$.
The dash-dotted curve is the SM including the resonance contributions
from the $\eta_c$ and the $\chi_{c0}$. Figure taken from \cite{HS}.}
\label{fig:q2}
\end{figure}

The resulting $B \to K \gamma \gamma$ di-photon mass spectra  
$d \Gamma/dq^2$ are shown in 
Fig.~\ref{fig:q2}. (For cuts and input, see \cite{HS}).
The SD part of the SM spectrum (solid curve) is completely hidden behind 
the $\eta_c$ and the $\chi_{c0}$ resonance contribution (dash-dotted curve).
The integrated SD branching ratio in the SM with cuts is small,
$\triangle {\cal{B}}(B \to K \gamma \gamma)_{SM}^{OPE} \simeq 1 
\times 10^{-9}$, with  about 50 \% uncertainty from the renormalization scale.
The contribution
from 1PI SD diagrams only gives a  somewhat reduced branching ratio 
${\cal{B}}(B \to K \gamma \gamma)_{SM}^{OPE \, 1PI} \simeq 0.5
\times 10^{-9}$. Such small SD branching ratios for 
double radiative decays are 
not a complete surprise. In fact, being of the same order in $\alpha_{em}$ 
as semileptonic rare decays with branching ratios of few$\times 10^{-7}$ 
\cite{Ali:1999mm}, we expect with $C_2 \simeq 1.1$ and 
$C_7^{SM} \simeq -0.3$
\begin{eqnarray}
{\cal{B}}(B \to K \gamma \gamma) \sim \left[ 
\frac{|\kappa_c Q_u^2 C_2|^2}{|C_9|^2+|C_{10}|^2} ~~ \mbox{or} ~~ 
\frac{|C_7|^2}{|C_9|^2+|C_{10}|^2} \right] \times
{\cal{B}}(B \to K \ell^+ \ell^-) \simeq {\cal{O}}
(10^{-9}).\nn
\end{eqnarray} 
The double photon decays are
substantially suppressed with respect to the semileptonic ones, since
the di-lepton operators with
large coefficients $|C_{9,10}^{\,SM}| \simeq {\cal{O}}(4-5)$
are not contributing to the former.

To be comparable to the resonance contributions at least in some region of 
phase space the SD contribution has to be lifted by roughly one order of 
magnitude above the SM one. 
At the same time, such type of NP should not violate other data. 
A sizeable 
enhancement of the photon dipole coupling, $C_7$, is excluded
by data on $B \to X_s \gamma$, which forces  
$|C_7| \simeq |C_7^{SM}|$. 
More room for NP is in the 4-Fermi operators, which contribute at
 higher order to $b \to s \gamma$ decays.
A possibility are large
non-standard (pseudo)scalar couplings to taus 
${\cal{O}}_{S(P)}^{\,\tau} =\frac{\alpha_{em}}{4 \pi} 
\bar s R b \bar \tau (\gamma_5) \tau$.
{}From ${\cal{B}}(B_s \to \tau^+\tau^-) <5 \%$ \cite{Grossman:1996qj}
one obtains
$|C^{\, \tau}_{S(P)}| \lsim 700$,
which is also consistent with data on other rare decays 
such as $b \to s \gamma$ and $b \to s \ell^+\ell^-$, $\ell=e,\mu$. 
In particular, large 
couplings are allowed because FCNC decays into a 
tau pair are  essentially unconstrained to date.
In other words, decays such as $B_s \to \tau^+ \tau^-$ and 
$B \to K^{(*)}  \tau^+ \tau^-$ have sizable room for NP.
As can be seen from Fig.\ref{fig:q2}, the corresponding maximal SD 
$B \to K \gamma  \gamma$ spectrum with NP in the $\tau$-couplings leaks 
marginally outside the resonance background.

Having a close look at the exclusive $B \to K \gamma \gamma$ decays, 
using the HQET and the SCET formalisms, we found that the resulting SD 
branching ratio in the SM is small, order $10^{-9}$, in 
agreement with the expectations based on the semileptonic rare 
$B \to K \ell^+ \ell^-$ decays. We  
discussed further contributions,
including long-distance effects via 
$B \to \eta_c  K \to \gamma \gamma K$ and 
$B \to  \chi_{c0} K \to \gamma \gamma K$. Unfortunately, we 
found that the SM SD contribution is not accessible 
behind the resonance peaks. We have further 
explored the possibility of NP contributions in $B \to K \gamma \gamma$. In particular, enhanced FCNC couplings to taus can give via the 1PI-loop  
branching ratios up to $\sim 10^{-8}$.
The corresponding maximal SD di-photon spectrum is then for some range of 
$q^2$ above the resonance background. 
We conclude that it is difficult to test SD physics with exclusive 
$B \to K \gamma \gamma$ decays. 
 However, even strongly resonance polluted, this mode could be useful to
probe the photon helicity in $b\to s \gamma$ via interference 
of different resonance amplitudes, as pointed out recently in
 \cite{Knecht:2005sc}.

%
%

\end{document}